\documentclass[preprint,aps,prc,showpacs,nofootinbib]{revtex4}
\usepackage{epsfig}
\usepackage{graphicx,amsmath}

\newcommand\ba{\begin{eqnarray}}
\newcommand\ea{\end{eqnarray}}
\newcommand\nn{\nonumber}
\newcommand{\be}{\begin{equation}}
\newcommand{\ee}{\end{equation}}

\begin{document}
\title{The two photon exchange amplitude in $ep$ and $e\mu $ elastic scattering: a comparison.}

\author{E. A. Kuraev }
\altaffiliation{\it JINR-BLTP, 141980 Dubna, Moscow region, Russian
Federation}

\affiliation{\it DAPNIA/SPhN, CEA/Saclay, 91191 Gif-sur-Yvette
Cedex, France}

\author{E. Tomasi-Gustafsson}
\affiliation{\it DAPNIA/SPhN, CEA/Saclay, 91191 Gif-sur-Yvette
Cedex, France}
\email{etomasi@cea.fr}

\date{\today}

\begin{abstract}
In this note we give arguments in favor of the statement that the contribution of the box diagram calculated for electron muon elastic scattering can be considered an upper limit to electron proton scattering. As an exact QED calculation can be performed, this statement is useful for constraining model calculations involving the proton structure.
\end{abstract}

\maketitle

The problem of the two photon exchange amplitude (TPE) contribution to elastic
electron-proton scattering amplitude has been widely discussed in the past. This amplitude has in principle a complex nature. Experimentally its real part, more exactly the real part of the interference between one and two photon exchange, can be obtained from electron proton and positron proton scattering in the same kinematical conditions. A similar information in the annihilation channel 
(electron-positron annihilation into proton-antiproton and in the reversal process) can be obtained from the measurement of the forward-backward asymmetry in the angular distribution of one of the emitted particles in the reaction center of mass (CMS) system. 

Recently, a lot of attention was devoted to the two photon exchange amplitude (TPE) in electron proton elastic scattering as a possible solution to a discrepancy between polarized and unpolarized measurements devoted to the determination of the proton form factors \cite{Jo00}. 

The theoretical description of TPE amplitude is strongly model dependent, as it involves modeling the proton and of its excited states, but it is still possible to derive rigorous results and predict exact properties of the two photon box:  model independent statements based on symmetry properties of the strong and electromagnetic interaction have been suggested in in Ref. \cite{Re99,Re04}. It has been proved that, due to C-parity conservation, the amplitude for $e^++e^-\to p+\bar p$, taking into account the interference between one and two photon exchange, is an odd function of $\cos\theta$, where $\theta$ is the angle of the emitted proton in the CMS of the reaction. This is equivalent, in the scattering channel, to destroy the linearity of the   Rosenbluth fit, i.e., the (reduced) differential cross section as a function of $\epsilon= [1+2(1+\tau)\tan^2(\theta_e/2)]^{-1}$, where $\theta_e$ is the electron scattering angle, for a fixed value of the momentum transfer squared, $Q^2$, between the incident and the outgoing electron. This property must be satisfied by all model calculations.

A second possibility is to do an exact calculation of the box diagram, which is possible for electron electron and electron muon scattering, and in the crossed channel (i.e., replacing the proton with a lepton) \cite{china}, where the muon can be considered a structureless proton. Even if such calculation is not rigorous when applied to the interaction on proton, the interest of a pure QED calculation is that the results should be considered as an upper limit for any calculation involving protons, as it will be discussed in this work.

The discussion of TPE box diagram in $ep$ scattering, can not be restricted to one proton intermediate state, but inelastic amplitudes should be consistently taken into account. Concerning the real part, the contribution to the amplitude from the proton on one side and from the inelastic intermediate states on the other side, are not gauge invariant, if considered separately. Only there sum is gauge invariant: the Ward-Takahashi identities relate the vertex function and the nucleon Green
function \cite{Kr66}. On the contrary, for the imaginary part, these contributions are separately gauge invariant, as the intermediate nucleon is on shell, as well as the external nucleons, therefore they must have comparable values.

Analyticity arguments lead to a (almost complete) compensation of
elastic and inelastic contributions in the whole amplitude. This statement is rigorous in QED \cite{Ba81}, and has been recently extended to electron-hadron scattering at small scattering
angles. Moreover,based on such statement, sum rules which relate peripheral cross section and elastic form factors have been derived in QCD and their validity verified on experimental data (see Ref. \cite{Ku06a} and Refs. therein).
Therefore, one can state that elastic and inelastic contributions are of the same order of magnitude, which is sufficient for our aim here. 

The notion of "nucleon form factors (FF)" can not be applied to the 
two-photon exchange amplitude (TPE) since one of the nucleons is off mass shell. Nevertheless the s-channel imaginary part of TPE, which corresponds to a single on mass shell nucleon and on mass shell electron in the intermediate state, can be analyzed in terms of FFs. Moreover it provides the gauge-invariant contribution to the imaginary part of the whole TPE. We build a simple model, calculating the $e\mu$ box Feynman diagram with one muon (nucleon) in the intermediate state. For the proton case, the muon mass is taken equal to the proton mass, and the proton structure is described by form factors. 

We can neglect the spin dependence and we calculate scalar four-dimensional integrals with point-like particles (in case of $e\mu$ scattering), and including proton form factors (for $ep$ scattering). A complete calculation was performed in Ref. \cite{Ni71} where similar scalar Feynman integrals with three and four denominators are involved. 

\begin{figure}
\begin{center}
\includegraphics[width=12cm]{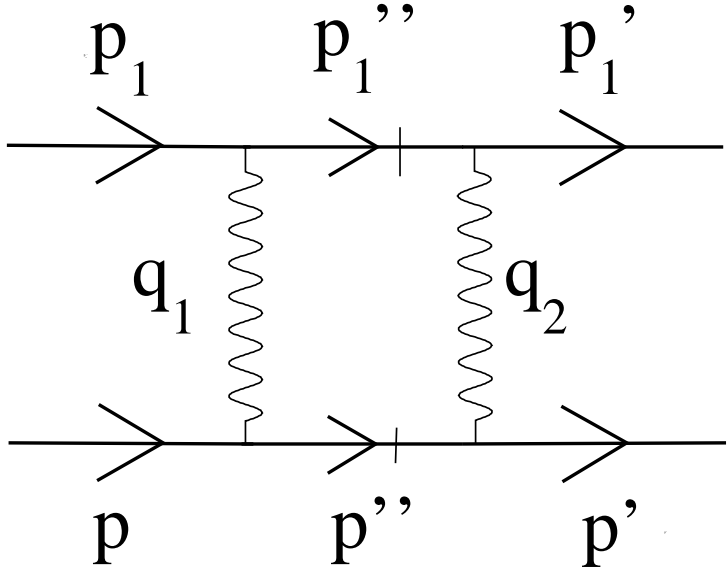}
\caption{\label{Fig:fig1} $s$-channel discontinuity of the Feynman box amplitude for $e\mu$ scattering.}
\end{center}
\end{figure}

Our aim is not to do a complete calculation of the box diagram, but to find an upper limit of this term: in every step, one should compare the relevant integrals. The purpose of this note is to prove that, modeling the proton by $Q^2$ decreasing form factors leads to a smaller contribution of the box diagram, compared to the QED case. We will prove this statement for the imaginary part of the amplitude corresponding to the box diagram
with one proton line connecting two $\gamma p p$ vertices and the validity for the relevant part of the full amplitude, ${\cal A}$, can be inferred through dispersion relations:
\be
{\cal A}(s,t)=\frac{1}{\pi}\int\frac {ds'Im {\cal A}(s',t)}{(s'-s-i\epsilon)}.
\label{eq:eq1}
\ee
Let us consider the cases where the target $T$ is a proton or a muon (Fig. \ref{Fig:fig1}) with the following convention for the particle four momenta:
\be
e(p_1)+ T (p)\to e(p_1'')+T (p'')\to e(p_1')+T (p').
\label{eq:eq2}
\ee

The following kinematical relations hold in the center of mass frame:

$p_1+p=p_1'+p'$, $q=p_1-p_1'$ ,~$p_1'+p'=p_1''+p''$,~$q_1=p_1-p_1''$, $q_2=p_1''-p_1',$

$Q_1^2=-q_1^2=-(p_1-p_1'')^2=2(\vec p)^2 (1-c_1),$

$Q_2^2=-q_2^2=-(p_1''-p_1')^2=2(\vec p)^2 (1-c_2),$

$Q^2=-q^2=-(p_1-p_1'')^2=2(\vec p)^2 (1-c),$

where $c_1= \cos\theta_1$ , $c_2= \cos\theta_2$, $c= \cos\theta$, and 
$\theta_1=\widehat{\vec p_1 \vec p_1''}$, $\theta_2=\widehat{\vec p_1''\vec p_1'}$, and $\theta=\widehat{\vec p_1 \vec p_1'}$. The momenta carried by the virtual photons are $q_1=k$ and $q_2=q-k$.

The contribution to the Feynman amplitude corresponding to the diagram of Fig. \ref{Fig:fig1} can be written as 
\be
{\cal M}=\frac{1}{(2\pi)^2}\int \frac{{\cal N} d\Gamma}{ (Q_1^2+\lambda ^2) (Q_2^2+\lambda ^2)}, 
\label{eq:eqm}
\ee
where ${\cal N}=F(Q_1^2)F(Q_2^2)$(${\cal N}=1$) for $ep$($e\mu$) scattering, $\lambda$ is a fictitious photon mass and $d\Gamma$ is the phase volume of the loop intermediate state. 
Taking into account the fact that the intermediate particles are on shell, 
one can write for the proton case:
\ba
d\Gamma &=&d^4p_1''\delta (p_1''^2-m^2)\delta (p''^2-M^2) d^4 p''\delta^4(p_1+p-p_1''-p'')\nn \\
&=&\frac{d^3p_1''}{2\epsilon_1''}\frac{d^3p''}{2\epsilon ''}\delta^4(p_1+p-p_1''-p'')
=\frac{d^3p_1''}{4\epsilon_1''\epsilon ''}\delta (\sqrt{s}-\epsilon_1''-\epsilon ''),
\nn \\
&&\epsilon_1'' =\frac{s-M^2}{2\sqrt{s}},~\epsilon'' =\frac{s+M^2}{2\sqrt{s}}\label{eq:eq3}.
\ea
Finally one can write
\be 
d\Gamma= \frac{s-M^2}{8s} dO_1'', 
\label{eq:eq4}
\ee
where $dO_1''$ is the solid angle of the electron in the intermediate state, which can be expressed as a function of the angles defined above as:
\be
dO_1''=\frac{2 dQ_1^2dQ_2^2}{\sqrt{{\cal D}_1Q_0^2}},~{\cal D}_1=2(Q_1^2+Q_2^2)Q^2Q_0^2 -2 Q^2Q_1^2Q_2^2 -(Q_1^2-Q_2^2)^2Q_0^2-(Q^2)^2Q_0^2,
\label{eq:eq5}
\ee
with the relation $Q_0^2=2\vec p^2=(s-M^2)^2/(2s)$. The positivity of the function ${\cal D}$ defines the solid angle kinematically available for the reaction.

Therefore one can write the contributions corresponding to the 'QED' diagram in Fig. \ref{Fig:fig1}, in case of a muon target:
\be
{\cal M}_{\mu}= \frac{1}{\sqrt{8s}} \int \frac{dQ_1 ^2 dQ_2^2}{\sqrt{{\cal D}_1} (Q_1^2+\lambda ^2) (Q_2^2+\lambda ^2)}.
\label{eq:eq6}
\ee
Introducing a generalized form factor for the proton, one finds for the 'QCD' diagram of Fig. \ref{Fig:fig1}, in case of a proton target:
\be
{\cal M}_{p}=\frac{1}{\sqrt{8s}} \int \frac{dQ_1^2 d Q_2^2F(Q_1^2) F(Q_2^2) }{\sqrt{{\cal D}_1} (Q_1^2+\lambda ^2) (Q_2^2+\lambda ^2)}. 
\label{eq:eq7}
\ee
Therefore the condition $F(Q_1^2) F(Q_2^2)<1$ is equivalent to the statement that the value of the electron-muon scattering amplitude can be considered an upper estimation of the amplitude for electron-proton scattering. 

Nucleon form factors are functions which are rapidly decreasing with $Q^2$. The Pauli and Dirac form factors, $F_1$ and $F_2$, are related to the Sachs form factors by :
\be
F_1(Q^2)=\frac{\tau G_M(Q^2)+G_E(Q^2)}{\tau+1},~ F_2(Q^2)=\frac{G_M(Q^2)-G_E(Q^2)}{\tau+1},~\tau= \frac{Q^2}{4M^2}, 
\label{eq:eq8}
\ee
with the following normalization: $F_1(0)=1$, $F_2(0)=\mu_p-1=1.79$, where $\mu_p$ is the magnetic moment of the proton in units of Born magneton.

\begin{figure}
\begin{center}
\includegraphics[width=12cm]{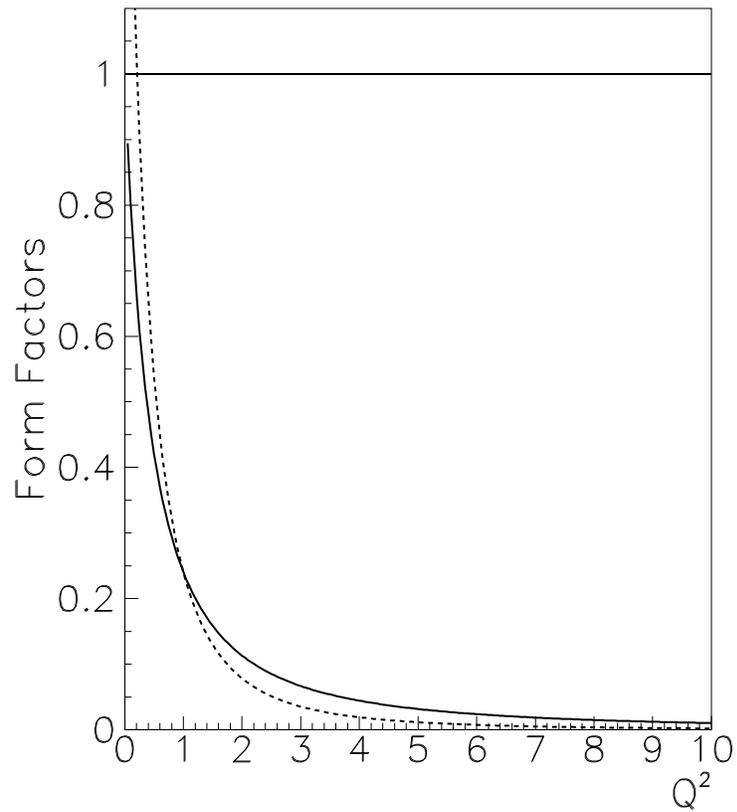}
\caption{\label{Fig:fig2}Form factors as a function of $Q^2$: $F_1(Q^2)$ (solid line), $F_2(Q^2)$ (dashed line).}
\end{center}
\end{figure}

\begin{figure}
\begin{center}
\includegraphics[width=12cm]{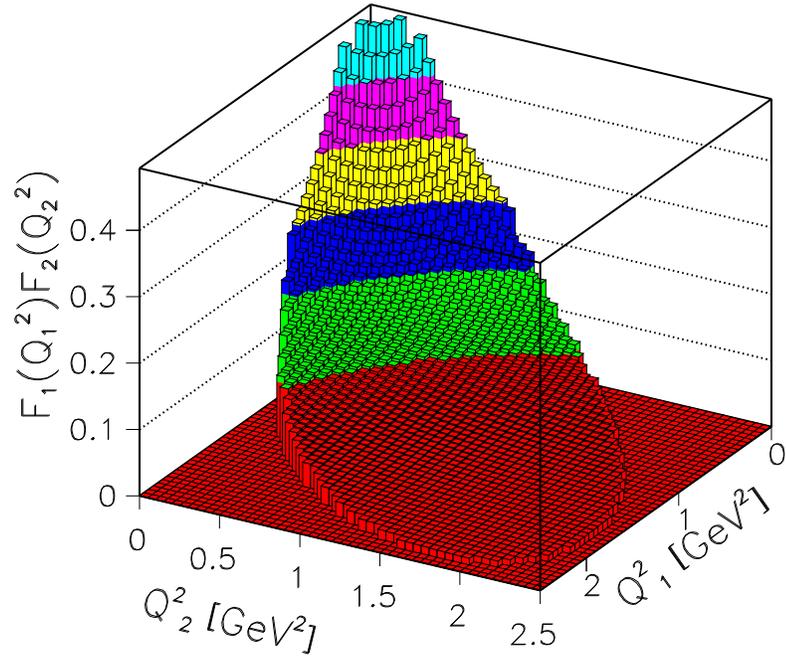}
\caption{\label{Fig:fig3}Bidimensional plot of $F_1(Q_1^2)F_1(Q_2^2)$ as function of $Q_1^2$ and $Q_2^2$. }
\end{center}
\end{figure}

\begin{figure}
\begin{center}
\includegraphics[width=12cm]{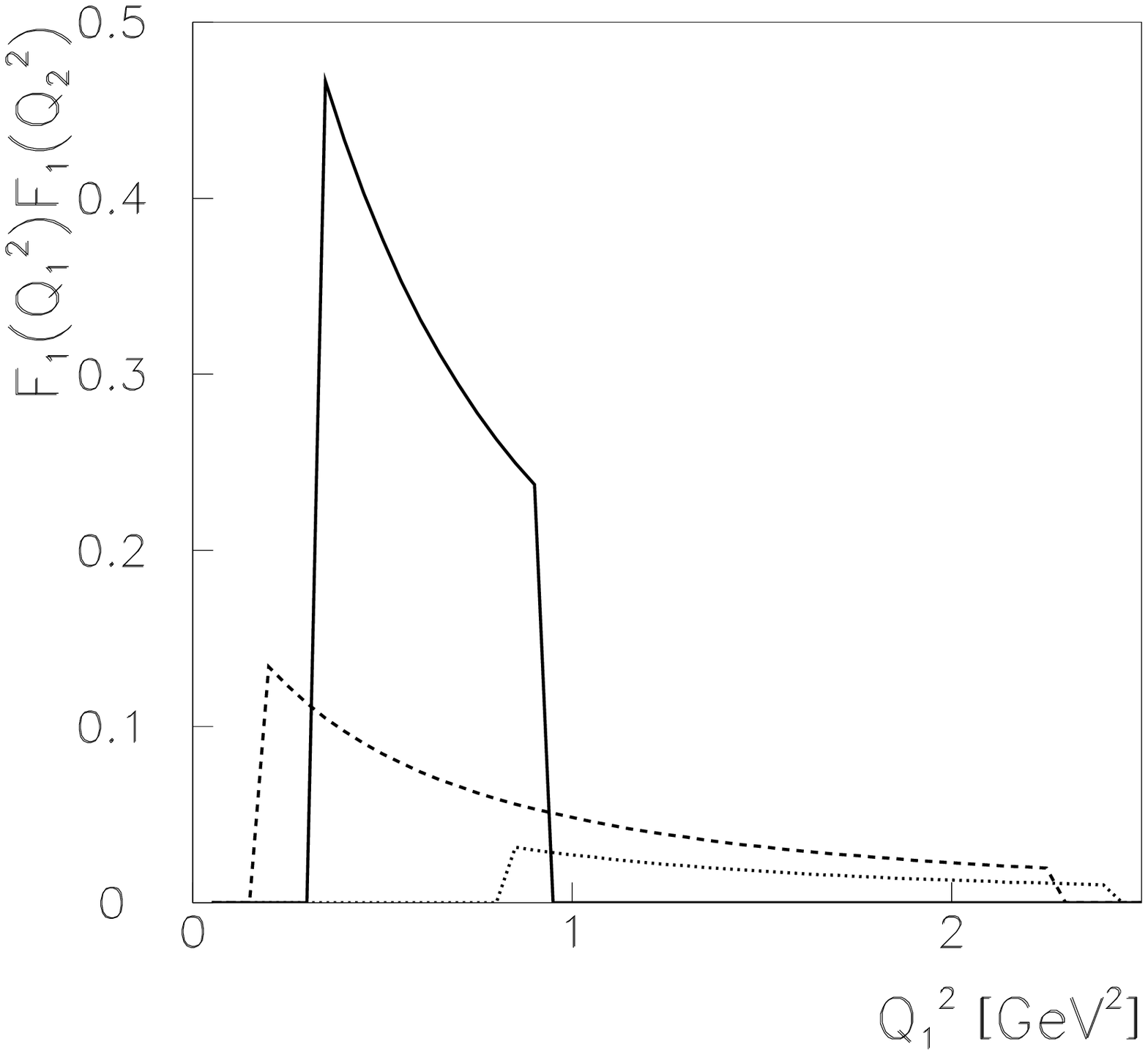}
\caption{\label{Fig:fig4}Projection on $F_1(Q_1^2)F_1(Q_2^2)$ on the $Q_1^2$ axis for $Q_2^2=0.05 $ GeV$^2$ (solid line), $Q_2^2=1.2 $ GeV$^2$ (dashed line),
$Q_2^2=2 $ GeV$^2$ (dotted line).}
\end{center}
\end{figure}

Let us consider the dipole approximation as a good approximation at least for the magnetic proton form factor $G_M$, although it has been shown that the electric form factor $G_E$ deviates from the dipole form \cite{Jo00}. In any case, any parametrization closer to the data will give even lower values as compared to the dipole form. In this approximation, we have:
\be
F_1^D(Q^2)=\frac{(\tau \mu_p+1)G_D(Q^2)}{\tau+1};~ F_2^D(Q^2)=\frac{( \mu_p-1) G_D(Q^2)}{\tau+1},~G_D(Q^2)=[1+Q^2\mbox{(GeV)}^2/0.71]^{-2}. 
\label{eq:eq9}
\ee
In Fig. \ref{Fig:fig2} we show $F_1(Q^2)$ (solid line), $F_2(Q^2)$ (dashed line), which are smaller than unity practically overall the $Q^2$ range. The product $F_1(Q_1^2)F_1(Q_2^2)$ is shown in Fig. \ref{Fig:fig3} as a bidimensional plot, and in Fig. \ref{Fig:fig4}, as a projection on the $Q_1^2$ axis for $Q_2^2=0.05 $ GeV$^2$ (solid line), $Q_2^2=1.2 $ GeV$^2$ (dashed line),
$Q_2^2=2 $ GeV$^2$ (dotted line).

One can see that the condition $F(Q_1^2) F(Q_2^2)<1$ is satisfied, starting from very low values of $Q^2$.  Let us stress that  $F_1(Q^2)$ is normalized to 1 and decreases with $Q^2$, being therefore smaller than unity; in the expression of the hadronic current, $F_2(Q^2)$ is multiplied by $q_{\mu}$, which lowers  its contribution at small $Q^2$, whereas at larger $Q^2$ it does not compensate the steep $Q^{-6}$ behavior of this form factor, as expected from quark counting rules \cite{Ma73}.

In conclusion we note that a destructive interference of the contributions of the single proton and its excited states takes place in (\ref{eq:eq1}), which results in additional suppression of the $ep$ amplitude compared to $e\mu$. A mutual compensation of the amplitudes for 'elastic' proton intermediate state with the excited hadronic states exists, and the reason lies in the superconvergent character of the dispersion relation (\ref{eq:eq1}), where the total amplitude is implied. Indeed, considering the amplitude for virtual Compton scattering in the complex $s$ plane,  closing the integration countour to the right hand singularities (which correspond to the proton intermediate state (pole) and to excited hadron states (cuts)) a compensation takes place, up to the small contribution of the left hand cut. Details are given in \cite{Ku06a}.

Therefore all model calculations for $ep$ elastic scattering as \cite{twogamma} should result in smaller contribution of the two photon amplitude, as compared to QED calculations \cite{china}.



\begin{thebibliography}{99}

\bibitem{Jo00}
M. K. Jones {\it et al.}, Phys. Rev. Lett. 84 (2000) 1398;
O. Gayou {\it et al.}, Phys. Rev. Lett. 88 (2002) 092301;
  V.~Punjabi {\it et al.},
  Phys.\ Rev.\ C 71 (2005) 055202
  [Erratum-ibid.\ C {\bf 71} (2005) 069902].

\bibitem{Re99} M. P. Rekalo, E. Tomasi-Gustafsson and D. Prout, Phys. Rev.
C {\bf 60}, 042202(R) (1999).

\bibitem{Re04}
  M.~P.~Rekalo and E.~Tomasi-Gustafsson,
  Eur.\ Phys.\ J.\ A {\bf 22}, 331 (2004);
  M.~P.~Rekalo and E.~Tomasi-Gustafsson,
  Nucl.\ Phys.\ A {\bf 740}, 271 (2004);
  M.~P.~Rekalo and E.~Tomasi-Gustafsson,
  Nucl.\ Phys.\ A {\bf 742}, 322 (2004).

\bibitem{china}
  E.~A.~Kuraev, V.~V.~Bytev, Y.~M.~Bystritskiy and E.~Tomasi-Gustafsson,
  Phys.\ Rev.\ D {\bf 74}, 013003 (2006).
\bibitem{Kr66}
N. M. Kroll, Il Nuovo Cimento XLV A {\bf 65} (1966).
\bibitem{Ba81}
V.~N.~Baier, E.~A.~Kuraev, V.~S.~Fadin and V.~A.~Khoze,
Phys.\ Rept.\  {\bf 78}, 293 (1981);
\bibitem{Ku06a}
  E.~A.~Kuraev, M.~Secansky and E.~Tomasi-Gustafsson,
  Phys.\ Rev.\  D {\bf 73}, 125016 (2006) and refs therein;
V.~N.~Baier, E.~A.~Kuraev, V.~S.~Fadin and V.~A.~Khoze,
Phys.\ Rept.\  {\bf 78}, 293 (1981);
  E.~A.~Kuraev, S.~Bakmaev, V.~V.~Bytev and E.~Tomasi-Gustafsson,
  arXiv:0711.3192 [hep-ph];
  E.~A.~Kuraev, V.~V.~Bytev, S.~Bakmaev and E.~Tomasi-Gustafsson,
  arXiv:0710.3699 [hep-ph].
\bibitem{Ni71}  
P. Van Nieuwenhuizen,  
Nucl.\  Phys.\   B {\bf 28}, 429 (1971).
\bibitem{twogamma}
P.~G.~Blunden, W.~Melnitchouk and J.~A.~Tjon,
Phys.\ Rev.\ Lett.\  {\bf 91}, 142304 (2003);
Y.-C. Chen, A. Afanasev, S. J. Brodsky, C. E. Carlson, and  M. Vanderhaeghen,
Phys. Rev. Lett.  93 (2004) 122301;
  P.~A.~M.~Guichon and M.~Vanderhaeghen,
  Phys.\ Rev.\ Lett.\  {\bf 91}, 142303 (2003);
  D.~Borisyuk and A.~Kobushkin,
  Phys.\ Rev.\  C {\bf 74}, 065203 (2006).
\bibitem{Ma73}
  V.~A.~Matveev, R.~M.~Muradian and A.~N.~Tavkhelidze,
  Lett.\ Nuovo Cim.\  {\bf 7}, 719 (1973); 
  S.~J.~Brodsky and G.~R.~Farrar,
  Phys.\ Rev.\ Lett.\  {\bf 31}, 1153 (1973).



\end{thebibliography}
\end{document}